# Comparison of user interfaces for measuring the matrix sentence test on a smartphone

Samira Saak[1,2*], Angelika Kothe[1,2], Mareike Buhl[2,3], Birger Kollmeier[1,2]

[1]Medical Physics, Department of Medical Physics and Acoustics, Carl von Ossietzky Universität Oldenburg, Oldenburg, Germany
[2]Cluster of Excellence Hearing4all, Carl von Ossietzky Universität Oldenburg, Germany
[3]Institut Pasteur, Université Paris Cité, Inserm, Institut de l'Audition, F-75012 Paris, France

**Abstract**

Using smartphones for mobile self-testing could provide easy access to speech intelligibility testing for a large proportion of the world population. The matrix sentence test (MST) is an ideal candidate in this context, as it is a repeatable and accurate speech test currently available in 20 languages. In clinical practice, an experimenter uses professional audiological equipment and supervises the MST, which is infeasible for smartphone-based self-testing. Therefore, it is crucial to investigate the feasibility of self-conducting the MST on a smartphone, given its restricted screen size.

We compared the traditional closed matrix user interface, displaying all 50 words of the MST in a 10x5 matrix, and three alternative, newly-developed interfaces (slide, type, wheel) regarding SRT consistency, user preference, and completion time, across younger normal hearing ($N$=15) and older hearing impaired participants ($N$=14).

The slide interface is most suitable for mobile implementation. While the traditional matrix interface works well for most participants, not every participant could perform the task with this interface. The newly-introduced slide interface could serve as a plausible alternative on the small screen of a smartphone. This might be more attractive for elderly patients that may exhibit more tactile and visual impairments than our test subjects employed here.

* Corresponding author. E-mail address: samira.kristina.saak@uol.de

Comparison of UIs for smartphone-based MSTs

**Introduction**

Using smartphones for mobile self-testing could provide easy access to speech intelligibility testing for a large proportion of the world population - especially for communities underserved with audiological practitioners. For this purpose, the matrix sentence test (MST) is an ideal candidate, as it is a repeatable speech test available in 20 languages covering 60 % of the world population (Hörzentrum Oldenburg gGmbH, n.d.-a; Kollmeier et al., 2015). The MST is used in research, for hearing aid benefit assessment, and slowly starting to be used in clinical contexts in a number of countries. As it tests sentence recognition in noise, it provides an ecologically more relevant estimate of individual hearing impairment in real life than a speech test in quiet, while being less dependent on a quiet test environment and an exact calibration of the test equipment (Kollmeier et al., 2015). Moreover, the test outcomes are still grossly related to the average audiogram (Wardenga et al., 2018). A smartphone implementation, however, is currently missing to also allow for individual self-testing and early detection of hearing deficits. If listeners conduct the MST by themselves (without experimenter), the current interface of the MST presents all 50 words at once to the user in form of a 10x5 matrix, which may prove problematic for a smartphone implementation, given the restricted size and input modalities of smartphones. This is especially relevant, as the majority of potential users of such a mobile implementation are above 65 years and could be affected by age-dependent declines in visual acuity, motor skills, and cognitive abilities (Farage et al., 2012; Salman et al., 2023; Wong et al., 2010).

First, a decline in visual acuity can lead to difficulties with detecting and discriminating details (Farage et al., 2012) on mobile interfaces/small screens. Second, age-related fine motor control declines reduce the precision of arm, hand, and finger movements (Hackel et al., 1992) and in turn, also increase the required time for task completion (Seidler et al., 2010). Presenting all 50 words of the MST could, therefore, be problematic on the small screen of smartphones, as small fonts might pose a problem, buttons, and button spacing might be too small (Hwangbo et al., 2013; Lee & Kuo, 2007), and too much information might be presented at once (Wong et al., 2010).Third, an age-related decline in cognitive abilities can result in reduced working memory capacities, and processing speed, among others (Park & Schwarz, 2000). The former could affect the ability to remember presented sentences, in that way adversely affecting task performance; the latter could more generally increase completion time and perceived task demands (Salthouse, 1996).

For a successful smartphone-based implementation of the MST, it is therefore crucial to define an interface that users can perform the tasks with and considers potential difficulties that elderly may experience with the application. In other words, a potential interface for the MST needs to be usable by the elderly target group. Lewis (Lewis, 2014) described *summative* usability as the ability to use a product for its intended purpose, to use it efficiently, and also with a feeling of satisfaction. Unfortunately, older users are often not considered in the development of technology applications (Chun & Patterson, 2012). As a result, usability for the elderly is often not optimal, fostering technological anxiety and inhibiting its uptake by the elderly (Frishammar et al., 2023).

To reduce the cognitive complexity for elderly with cognitive impairments, the simplified MST could be used. The simplified MST is a reduced version of the MST, only contains 3-word sentences, e.g. "seven old windows", and is currently available in 5 languages (Hörzentrum Oldenburg gGmbH, n.d.-b; Wagener & Kollmeier, 2005). The German version of the simplified MST has been termed OLKISA since it was first intended to be used with children, but later turned out to be useful for elderly patients as well. It could prove to be a useful speech test for smartphones, given that it reduces the cognitive workload of the task, the amount of information that needs to be displayed on the smartphone, and the number of physical interactions with the interface.

Comparison of UIs for smartphone-based MSTs

Next to fulfilling usability requirements for the elderly, an application needs to be usable within its environment. If laboratory or advanced audio equipment are required, most individuals will not have access to easy self-testing. In contrast, readily available equipment may increase the uptake of a proposed mobile speech test. In that direction, the feasibility of audiometric screening via smartphones with inexpensive headphones has been shown (Hussein et al., 2016; Swanepoel et al., 2014). Likewise, the Digits-in-Noise test has proven successful as a screening test via telephone and internet (Smits et al., 2004; Zokoll et al., 2012). The speech test consists of spoken numbers, which can be selected from a telephone interface. While this provides a test that is easy to conduct, it also comes with a disadvantage: Numbers do not represent every-day-language and its speech spectrum as appropriately as syntactically correct sentences. The MST speech material was optimized in several regards, for example to represent the phoneme distribution or the frequency of occurrence of different words of the respective language and to provide a high testing efficiency in terms of decrease in measurement variability per unit of measurement time (Kollmeier et al., 2015). It is, thus, more representative for every-day life and could provide a more detailed and precise assessment for a given amount of measurement time than the digits-in-noise test, used for screening.

Several implementations already exist for measuring the German MST (Oldenburg Sentence Test, OLSA) outside the lab. Ooster et al. (Ooster et al., 2020) implemented a system for measuring the OLSA via Amazon Alexa using automatic speech recognition (ASR). Likewise, Bruns et al. (Bruns et al., 2022) developed a voice-over-IP system for measuring the OLSA via telephone, also using an ASR system. Both implementations come with several advantages and disadvantages. First, both systems are based on ASR systems. While ASR systems can ease the usability of an application, the training of ASR systems across multiple languages is, unfortunately, a non-trivial task. In contrast, a closed mobile interface for the OLSA could directly be applied in all 20 languages currently available – and easily be adapted to later included languages. Second, owning Amazon ALEXA is not as common as a telephone, or a smartphone device (between 2014-2018 about 70 million Amazon ALEXAs were sold, while 6,970 million smartphone subscriptions were reported (Ericsson, 2023; Laricchia, 2022). In that way, the number of potential users is reduced with Amazon ALEXA or comparable ASR-controlled service provision units.

Implementing the MST on a smartphone and providing an appropriate user interface (UI) is, therefore, highly relevant to increase the group of potential users, whose hearing can be characterized. By addressing smartphone users and providing a tool easily available across multiple languages this can be achieved. It is, however, not yet known to what extent elderly hearing impaired (oHI) individuals can conduct the MST and simplified MST accurately on a smartphone, given the combination of a small screen size, and potential age-dependent declines in cognitive ability, visual acuity, and motor skills. With the present study, therefore, we aim at (1) testing the general feasibility of measuring the MST on a smartphone. For that purpose, we investigate the performance of younger normal hearing (yNH) individuals on smartphone implementations of both the German MST (OLSA) and the German simplified MST (OLKISA). We hypothesize that this will be feasible with a smartphone and household in-ear headphones. We further aim at (2) characterizing any specific needs regarding such a smartphone implementation that oHI individuals might have. To that end, we compare the yNH and the oHI group with respect to usability aspects, namely, their performance consistency, completion time and user preference, across four suggested potential interfaces for the OLSA, and the typical (matrix) interface of the OLKISA. The four interfaces differ regarding layout and response options. Ultimately, we (3) propose an interface for the OLSA that both elderly and young users can easily cope with, is time efficient, and provides accurate SRT results. Hence, our research should allow us to answer the following research questions:

**RQ1:** Is it feasible to obtain the same OLSA and OLKISA results as in the lab with a calibrated smartphone and inexpensive household headphones with both yNH and oHI individuals?

**RQ2a:** Is there a group effect (age and hearing impairment; yNH vs. oHI) on SRT consistency, completion time, and user preference?

**RQ2b:** If a group effect is present, is it specific to certain interfaces?

**RQ3:** Which OLSA interface is most suitable for both oHI and yNH in terms of SRT consistency, time efficiency, and user preference?

**Materials and Method**

*Participants*

We recruited 32 participants for our study: one yNH group, and one oHI group with mild to moderate hearing impairment. The yNH group was recruited via posts at the University website, while participants for the oHI group were recruited via the Hörzentrum Oldenburg gGmbH. All participants gave written informed consent and were paid for their participation in the study. Inclusion criteria for the yNH group were an average PTA < 20 dB HL and age < 50 years; inclusion criteria for the oHI group were a symmetric hearing loss (PTA difference < 10 dB) with an PTA > 20 dB HL, and age ≥ 50 years. Participants were required to perform the smartphone measurements with their fingers. Fifteen participants qualified for the yNH group (*mean age* = 23.67, *SD* = 2.32, *female* = 66.7%) and 14 qualified for the oHI group (*mean age* = 73.77, *SD* = 6.73, *female* = 53.8%). The remaining three participants (oHI) were excluded, as they were not able to operate a smartphone sufficiently with their fingers, but instead required a pen. **Figure 1** depicts the audiogram ranges for the yNH and oHI groups.

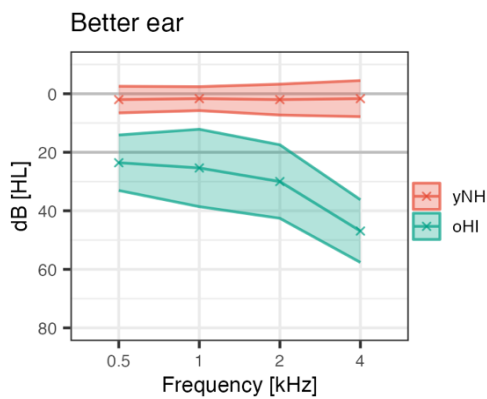

*Figure 1. Audiogram ranges for the yNH (red) and oHI (green) groups for their respective better ear. Means are shown as well as ranges corresponding to one standard deviation for yNH and oHI.*

*Research Design*

We employed a crossover study to examine the overall feasibility of measuring both the OLSA (Wagener, 2004) and OLKISA (Wagener & Kollmeier, 2005) on a smartphone, and to investigate which interface is best for measuring the OLSA. The measurement language was German. The OLSA consists of a 10x5 matrix in which sentences with fixed syntactical structure are built by selecting word combinations from the five name-verb-number-adjective-object columns. For instance, "Peter has seven old windows". The OLKISA is a reduced version of the OLSA, consisting of a 10x3 matrix (number-adjective-object). Each participant conducted several measurement conditions of the OLSA and the OLKISA in both a reference laboratory control session and in the smartphone test session. The order of test and control session, and the order of the measurement conditions were randomized. All tests were conducted in a soundproof-booth and both sessions used calibrated setups to control the levels and to allow comparing the influence of equipment and UIs. Ethical approval was obtained from the Research Ethical Committee of the Universität Oldenburg [Drs. 71/2015].

*Laboratory control session*

The laboratory control session was conducted using the software Oldenburg Measurement Applications (OMA) by the Hörzentrum Oldenburg gGmbH with HDA200 headphones. The OMA is a CE-certified medical product that can measure the audiogram, as well as the OLSA and OLKISA, among other tests. By using this software, the exact same measurements are performed as in clinical practice. Both the OLSA and OLKISA in OMA were measured in an open version to mimic the standard clinical procedure. That means, participants reported the understood words verbally to the experimenter (open version), instead of selecting the respective words from a matrix displayed on a screen (closed version). OMA uses the adaptive procedure by Brand and Kollmeier (Brand & Hohmann, 2002) to control levels of speech and noise. Calibration was performed using the test-specific noise of the OLSA and an artificial ear.

*Smartphone test session*

We developed a web-based implementation to measure both the matrix and the simplified matrix test via an interface optimized for smartphones. The underlying measurement procedures mimic the procedure used by OMA. The backend of the application was build using Python Flask, Octave, and Bash. Python Flask is a common framework for web-development. Octave was chosen for implementing the sentence tests to reuse available measurement scripts for the OLSA (see (Schädler, 2021)). The communication between Python Flask and Octave was then enabled via the shell language Bash. The frontend and the interfaces were built using the common mixture of JavaScript, HTML, and CSS. A Linux laptop served as a server to host the application. The website was then opened via a browser on a OnePlus Nord N10 (Android operating system) smartphone with inexpensive in-ear Headphones (Sony MDR-XB50AP). For data security reasons, the current website should only be hosted within a secure network. Hence, we used eduroam. The adaptive procedure used fixed decreasing step sizes (+/- 10, 6, 5, 3, 2, 1, 1.5, 0.5). The smartphone and in-ear headphones were calibrated across frequencies (125 – 10000 Hz) using the KEMAR artificial head. The resulting noise level was also checked to match the control session.

*Interfaces for the OLSA & OLKISA*

To investigate how to best measure the OLSA on a smartphone, we developed and compared four potential interfaces (see **Figure 2**). The *matrix* interface serves as a reference interface and corresponds to the traditional closed interface of the OLSA. Here, the complete matrix is displayed at once. The *slide* interface displays the columns of the matrix sequentially. The next column is presented as soon as a word is selected, which makes the interface faster. The buttons and the font are larger than with the matrix interface, but the words have to be provided in the given order (name-verb-number-adjective-objective). The *type* interface requires users to type in the understood words in any order. Suggestions of words from the complete 50 words of the matrix are provided based on the already typed input. As such, this interface serves as a mixture between an open and a closed version of a MST. The *wheel* interface displays the columns as wheels that can be scrolled horizontally. All words can be selected in any given order and fonts and buttons are larger than with the matrix interface. The larger font and button sizes of the alternative interfaces aim to reduce the potential impact of visual and fine motor declines. Further, the *wheel* and *type* interface allow users to simultaneously see all selected words, whereas the *slide* interface immediately shows the next column. The *type* and *wheel* interface further assess different input modalities, namely horizontal scrolling vs. typing. The order of presented interfaces was randomized.

The OLKISA was measured only with the *matrix* interface, as we assumed that the screen size of smartphones is sufficient to display a 3x7 matrix of words.

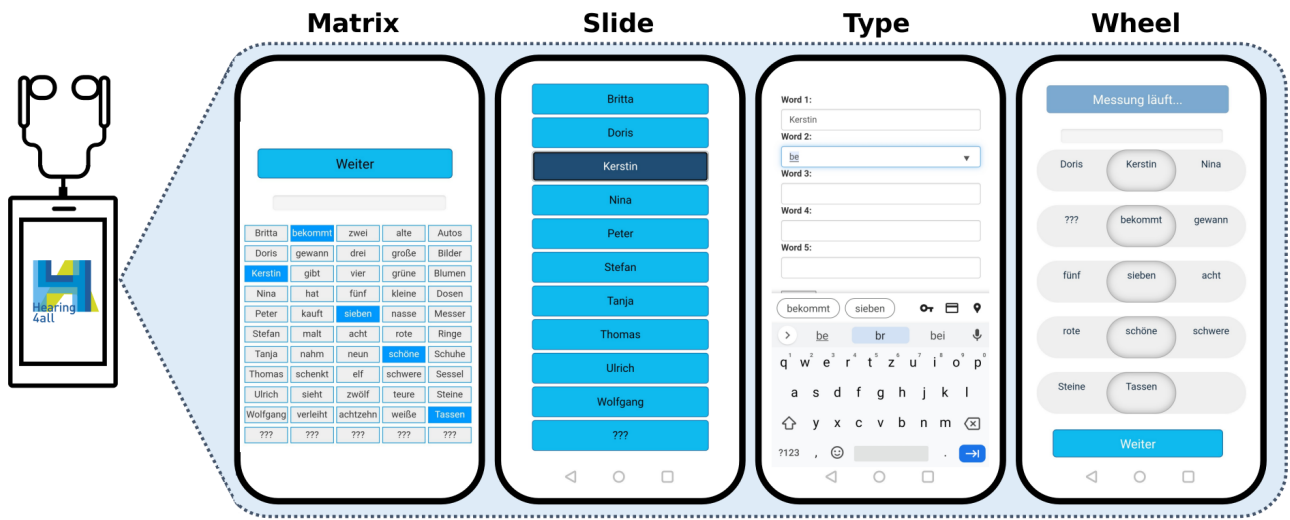

*Figure 2.* Matrix Sentence Test (MST) interfaces. The four interfaces under investigation. "Matrix" corresponds to the traditional closed MST UI. "Slide" presents the columns of the matrix sequentially. "Type" asks users to type in the words, while providing suggestions of words from the matrix. "Wheel" requires users to scroll the words horizontally.

*OLSA and OLKISA conditions*

The OLSA was measured with ICRA1 (S0N0, S0N90 binaural & monaural) and ICRA5 noises, i.e., one stationary and one fluctuating noise signal (Dreschler et al., 2001). During the measurement, the noise level was fixed at 65 dB SPL and the speech level was adapted starting from an SNR of 0 dB. The SRT50 was measured, i.e., the threshold where 50 % of the words are correctly understood. In the smartphone application, ICRA1 was measured across interfaces. ICRA1 S0N90 binaural & monaural, and ICRA5 were only measured with the slide interfaces to reduce testing time. We chose the slide interface for this, as we expected all participants to be able to conduct the measurements with this interface, whereas the small fonts and buttons of the matrix interface could prove difficult for the oHI group. For the spatial conditions (S0N90 binaural & monaural) the noise direction (90° from left or right) was chosen according to the worse ear of the participant, respectively. For the monaural condition no stimulus was presented to the better ear. Binaural measures were included as they provide more realistic testing situations. They provide relevant information regarding binaural and spatial hearing that cannot be captured with the collocated S0N0 condition (Pastusiak et al., 2019). The OLKISA was also measured with ICRA1 and ICRA5 in S0N0. The OLSA test list consisted of 20 sentences, whereas the OLKISA was measured with 14 sentences. For a tabular overview of all conditions see column "Condition" of **Table 2**)

Comparison of UIs for smartphone-based MSTs

*Questionnaire data*

Within the smartphone application, participants answered several short questions regarding age, gender, and general smartphone usage. In addition, users could provide open comments regarding the distinct interface. Finally, after completion of both the laboratory control and smartphone test sessions, participants were asked to rank the interfaces from highest to lowest preference, and could provide verbal comments regarding the different interfaces, which was noted by the experimenter.

***Procedure***

First, the pure-tone audiogram (500, 1000, 2000, 4000 Hz) was measured with OMA to control for the correct group allocation (yNH, oHI). Second, two training runs were conducted with the OLSA to counter potential training effects. The training lists were measured with the closed OMA version to familiarize the participants with the speech material. The first list was measured in quiet, the second list in the collocated S0N0 condition with test-specific noise (65 dB SPL). Next, participants were randomly allocated to either start with the laboratory control session (OMA), or with the smartphone test session (WEB). Within both sessions, the different conditions were measured in random order. The only exception was that the S0N0 ICRA1 slide condition was always measured prior to the remaining slide conditions. This was to ensure that the first encounter with each interface was measured in the same noise and spatial condition. Upon completion, the remaining session was measured. As a final step, participants were asked to rank the interface according to their preference. They also could provide verbal comments to the experimenter regarding the interfaces.

***Analyses***

*SRT consistency*

To test our hypothesis (RQ1) that both yNH and oHI can self-conduct the matrix and simplified matrix test, we compared paired SRT scores of participants for the smartphone test session to the laboratory control session. For this, we calculated both the root mean square error (RMSE) and bias for the different interface conditions (matrix, slide, type, wheel), spatial conditions (S0N90 monaural & binaural), and noise (ICRA1, ICRA5) conditions. For the interfaces, we further tested for significant differences to the laboratory control session ($p < 0.05$). To determine which interface is best in terms of SRT consistency (first part of RQ2), we tested for significant differences of the interfaces compared to the control condition across the two groups (yNH, oHI). To assess potential interface effects on SRT consistency, we further tested for interaction effects. For this, we used a linear mixed model nested within participants, with the following formula:

$$SRT = intercept\ (OMA) + UI + GROUP + UI*GROUP + (1\ |Participant)$$

Where *SRT* is the response variable, intercept refers to the control condition, *UI* refers to the specific user interface (matrix, slide, wheel, type), *GROUP* refers to either the yNH or the oHI group, *UI\*GROUP* tests for an interaction effect, and *(1 |Participant)* refers to the design of the model being nested within subjects.

*Completion time*

We analyzed the completion time of the four interfaces across the yNH and oHI groups via boxplots. This was done to find out which interface is fastest (second part of RQ2).

*User preferences*

To evaluate user preferences, we analyzed the provided rankings of the participants. To that end, we calculated the preference ratios for the interfaces to determine the overall preferred interface. That means, we calculated how often (in percent) a given interface was ranked higher than each of the remaining interfaces, for both the yNH and oHI group. As a result, we could compare pairwise preference differences between the two groups for each interface. Next, we analyzed the verbal or written comment data using conceptional concept analysis (Hsieh & Shannon, 2005). For each interface, written comments obtained by the app and verbal comments of the participants were combined. They were then categorized into different concepts in an inductive manner. This means, concepts were generated during the coding, and updated if new concepts were detected in the comments. To ensure all contained concepts were covered, this process was restarted several times. After a final set of concepts was determined, their occurrence across interfaces was counted. Concept analysis was performed in German, as participants provided their comments also in German. The concepts and examples were then translated to English via DeepL for an unbiased and standardized translation (see **Table 3** for some examples and **Supplementary Table 1** for all concepts with explanations).

**Results**

In the following we present the results of the SRT, completion time, and preference results for both the oHI and yNH group. In the yNH group every participant owned a smartphone. 73.33 % used it almost continuously during the day; 26.67 % used it once or more during the day. In contrast, for the oHI Group 10 % used it almost continuously during the day, and the remaining 90 % used it once or more a day.

*SRT consistency*

*OLSA interfaces*

**Figure 3** depicts the comparison of speech recognition thresholds (SRT) for the laboratory control session (OMA) and the smartphone test session (WEB). The yNH group was able to accurately measure the OLSA with a smartphone, as well as with all interfaces. The differences of the interfaces to the control session are not significant (p › 0.05, see **Table 1**) and the RMSE are around the test-retest error of 1 dB. The bias is slightly higher for the type and wheel interface than for the matrix (lowest) and the slide interface (**Table 2**).

The oHI group had the highest SRT consistency with the matrix interface, as indicated by (1) the low RMSE and bias scores (**Table 2**), and (2), the *UI* coefficient of the linear mixed model (**Table 1**). However, we excluded the visible outlier of the matrix interface (SRT = 5/-5 dB SNR) from the statistical analyses, as it would have strongly biased the analyses. This participant highlights that while most users would achieve accurate SRT scores with the matrix interface, some may not be capable of performing speech-in-noise tests on a smartphone with the matrix interface. A mobile implementation suitable for OLSA measurements should, however, be robust enough to avoid such outliers. For the other interfaces RMSE scores are higher, but also no extreme outliers are found. The differences between wheel and slide are marginal. While the wheel interface has slightly lower RMSE scores, it also results in slightly higher bias scores and *UI* coefficients. The type interface consistently had the highest SRT elevation, as evident from **Figure 3**, the RMSE results, and biases.

Comparison of UIs for smartphone-based MSTs

Overall, the matrix interface is most accurate for both the yNH and oHI, with the exception of the outlier. We can observe an effect of the oHI group on the performance with the different interfaces. The lowest interface-specific effects can be observed with the wheel interface, indicating that the wheel interface was most stable across the two groups (yNH and oHI).

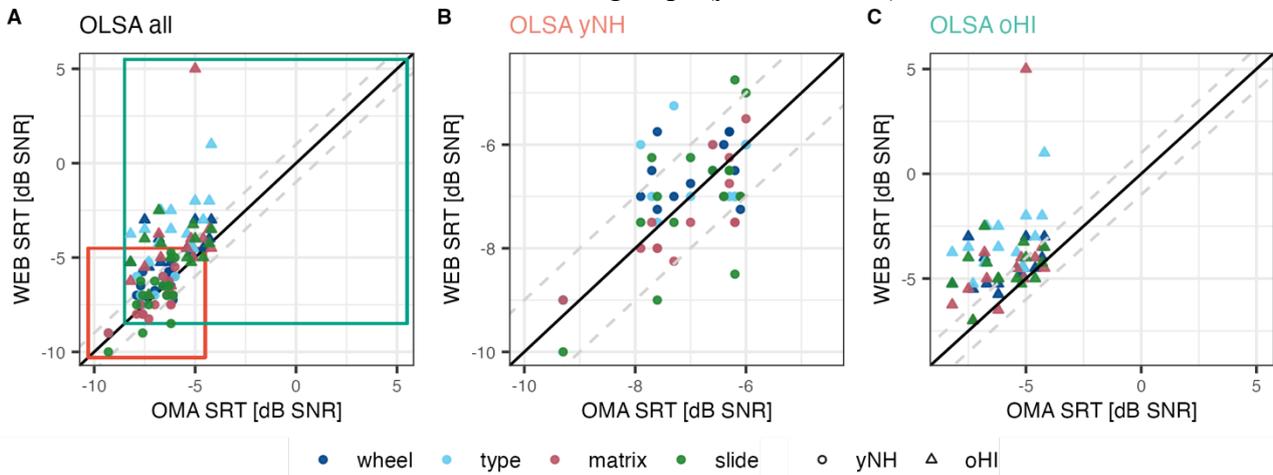

*Figure 3.* Comparison between SRTs obtained with the smartphone web-based interface (ordinate) vs the reference laboratory implementation using Oldenburg Measurement Applications (OMA, abscissa). (A) SRT results for the younger normal hearing (yNH) group (circles) and the older hearing impaired (oHI) group (triangles) across all four interfaces for ICRA1 in the S0N0 session (displayed in different colors). The dashed line corresponds to a 1 dB deviation from the line of perfect agreement. Data points within the dashed lines are within the expected test-retest error. (B) SRT results for the yNH group (zoom from red box in A) (C) SRT results for the oHI group (zoom from the green box in A).

*Table 1.* Main and interaction effects of two linear mixed models for UI on SRT score with either yNH or oHI as the reference category for the group (SRT = intercept (OMA) + UI + GROUP + UI*GROUP + (1|Participant). Interaction effects are only shown for the yNH reference model; for the oHI reference model the direction of the UI coefficients would be reversed. Laboratory control session (OMA) serves as the control (intercept). UI and group (yNH/oHI) are fixed effects nested within subjects. Bold values indicate that no significant difference to the intercept exists.

|  | **Coefficients** |  | **Estimate** | **Conf. Interval** | **P-value** |
|---|---|---|---|---|---|
| yNH | intercept | OMA | -6.967 | (-7.515,-6.418) | 0.0000 |
|  | UI | UImatrix | **-0.350** | **(-0.899,0.199)** | **0.2273** |
|  |  | UIslide | **-0.050** | **(-0.796,0.696)** | **0.8626** |
|  |  | UItype | **0.050** | **(-0.499,0.599)** | **0.8626** |
|  |  | UIwheel | **0.300** | **(-0.249,0.849)** | **0.3003** |
| oHI | intercept | OMA | -5.977 | (-6.612,-5.201) | 0.0000 |
|  | UI | UImatrix | 1.015 | (0.885,2.429) | 0.0014 |
|  |  | UIslide | 1.419 | (0.617,2.161) | 0.0000 |
|  |  | UItype | 2.746 | (1.992,3.536) | 0.0000 |
|  |  | UIwheel | 1.419 | (0.689,2.233) | 0.0005 |

Comparison of UIs for smartphone-based MSTs

| | | | | |
|---|---|---|---|---|
| GROUP | oHI | 0.990 | (0.184,1.795) | 0.0264 |
| UI*GROUP | matrix*oHI | 1.365 | (0.560,2.171) | 0.0017 |
| | slide*oHI | 1.469 | (0.664,2.275) | 0.0007 |
| | type*oHI | 2.696 | (1.890,3.502) | 0.0000 |
| | wheel*oHI | 1.119 | (0.314,1.925) | 0.0094 |

*OLSA spatial conditions and fluctuating noise (slide interface)*

The ICRA1 S0N90 binaural condition resulted in adequate test-retest values for the yNH group, while the SRTs with the ICRA1 S0N90 monaural condition yield an underestimation for the smartphone web-based version (**Figure 4**), most likely due to interaural crosstalk effects (see discussion). For the oHI group ($N = 11$ due to missing data) a slight bias can be observed in both conditions (see **Table 2**), similar to the bias observed with ICRA1 S0N0, but adequate consistencies between smartphone and laboratory UI are achieved. For the fluctuating noise condition ICRA5, we can, again, observe a bias similar to other slide interface conditions. The bias and RMSE are higher for the ICRA5 condition. Since the expected test-retest variability with ICRA5 is generally larger (2 dB for NH listeners, (Wagener, 2004)), this effect is in line with expectations.

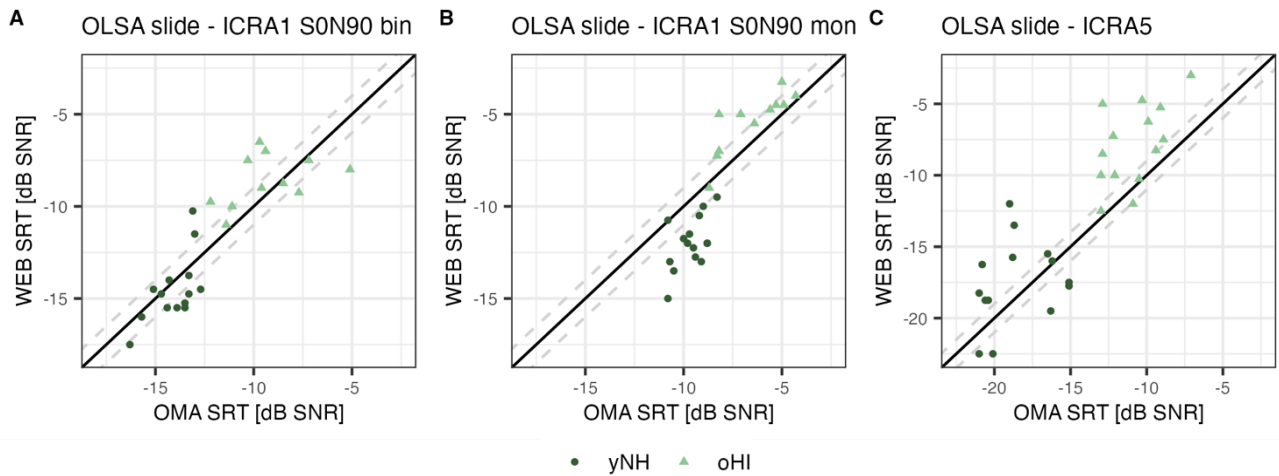

***Figure 4.*** *SRT results for slide condition. Dashed lines indicate expected test-retest variability (1 dB for ICRA1; 2 dB for ICRA5 (NH)).*

*OLKISA*

Due to the higher test-retest variability of the OLKISA (Wagener & Kollmeier, 2005), the comparison between smartphone UI vs. laboratory control UI session is expected to result in larger RMSE values than for the OLSA. Since the S0N0 ICRA5 condition was only measured using the slide interface, only the results for the respective slide interface conditions are displayed in **Figure 5** for the OLKISA and OLSA. For both the yNH and oHI groups the OLKISA resulted in higher RMSE (as expected) and bias than the OLSA with the matrix interface (**Table 2**). For ICRA5, we need to compare the OLKISA matrix interface to the OLSA slide interface, as ICRA5 was only measured with the slide interface for the OLSA. We again observe higher RMSE and bias values for OLKISA with the yNH group. In contrast, for the oHI group, we might observe an interface effect, as RMSE and bias scores

Comparison of UIs for smartphone-based MSTs

are slightly lower for OLKISA. However, this may be caused by the comparison of the slide and matrix interface, where the slide interface generally resulted in higher scores.

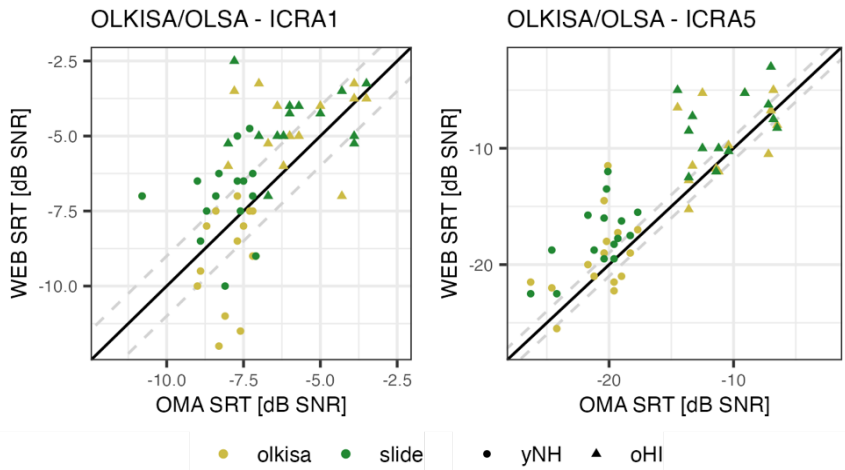

*Figure 5.* SRT results for the OLKISA and comparison to the OLSA slide interface for both the smartphone test session and the laboratory control session. Dashed lines indicate test-retest variability (1 dB for ICRA1; 2 dB for ICRA5).

*Table 2.* Condition overview, RMSE and bias for smartphone (Web SRT) vs. laboratory tests (OMA SRT) for all conditions across all participants, and separately for the yNH and oHI group.

| Condition | | | | RMSE (dB) | | | BIAS (dB) | | |
|---|---|---|---|---|---|---|---|---|---|
| **Test** | **Spatial condition** | **Noise** | **Interface** | **All** | **yNH** | **oHI** | **All** | **yNH** | **oHI** |
| Matrix sentence test (OLSA) | S0N0 | ICRA1 | matrix | 1.07 | 0.68 | 1.335 | 0.28 | -0.35 | 0.95 |
| | | | slide | 1.54 | 1.01 | 1.96 | 0.64 | -0.05 | 1.39 |
| | | | type | 2.24 | 0.91 | 3.08 | 1.36 | 0.05 | 2.76 |
| | | | wheel | 1.43 | 0.77 | 1.89 | 0.86 | 0.30 | 1.46 |
| | | ICRA5 | slide | 3.48 | 3.18 | 3.78 | 1.79 | 0.94 | 2.98 |
| | S0N90 binaural | ICRA1 | slide | 1.57 | 1.38 | 1.74 | 0.20 | -0.49 | 0.52 |
| | S0N90 monaural | ICRA1 | slide | 2.07 | 2.60 | 1.28 | -0.63 | -2.34 | 0.81 |
| Simplified Matrix sentence test (OLKISA) | S0N0 | ICRA1 | olkisa | 2.02 | 2.05 | 2.00 | -0.14 | -0.77 | 1.12 |
| | | ICRA5 | olkisa | 3.35 | 3.36 | 3.33 | 1.66 | 1.44 | 1.49 |

Comparison of UIs for smartphone-based MSTs

*Completion time*

**Figure 6** compares the absolute completion times for the different OLSA interfaces, as well as the OLKISA to infer completion times for practical implementation. The fastest test was the OLKISA. This was expected, as the OLKISA is a reduced version of the OLSA with only three words per sentence and 14 sentences in a test list. The matrix interface ranks second but is only marginally faster than the slide interface. The type interface took the longest time for both groups. The yNH group was generally faster in completing the test with all interfaces, as compared to the oHI group. This difference appears stable across all interfaces but the type interface. With the type interface, we can observe an interaction effect between the interface and the oHI group. That is, the oHI group took much longer in completing the test with this interface as compared to the remaining interfaces.

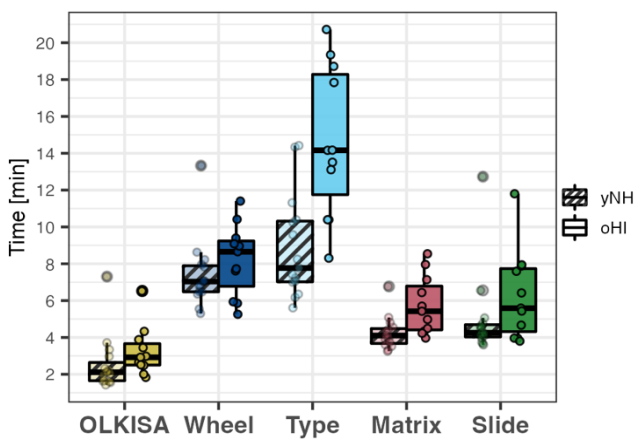

*Figure 6. Completion time of the OLSA (20 sentences) measured with ICRA1 S0N0 for each interface, as well as OLKISA (14 sentences), separate for yNH and oHI individuals.*

*Interface ranking*

**Figure 7** compares the interface rankings between oHI and yNH as expressed in the ratio how often an interface (color) was preferred over another interface (shape). The line indicates perfect agreement between yNH and oHI with respect to the ranking of a given interface. Symbols in the upper triangle indicate that oHI preferred this interface over another interface to a greater extent than yNH. Overall, between yNH and oHI only slight preference differences exist. The results indicate that on average the matrix interface (red symbols) was mostly preferred by both yNH and oHI, especially since the matrix UI was preferred in the direct comparison between slide and matrix UI by both groups. However, the matrix interface was generally more strongly preferred by the yNH. This becomes most visible by the matrix-wheel comparison (red square). The yNH group preferred the matrix interface over the wheel interface to a greater extent (red square). The resulting higher preference of the oHI for the wheel interface also transfers to the comparison with the type interface. Here, we can observe a slight preference of the oHI for the wheel interface, and for the yNH for the type interface (blue circle, turquoise square).

The second highest ranking was achieved by the slide interface (green symbols). In comparison to the matrix interface, it is only slightly less preferred (green triangle, red diamond) while the yNH group showed a stronger relative preference for the matrix UI than the OHI group. Further, the slide UI consistently ranked higher than the type and wheel interface. Noteworthy is that for the oHI group

Comparison of UIs for smartphone-based MSTs

the slide interface had a higher ranking than the matrix interface in comparison to the wheel interface (green square vs. red square).

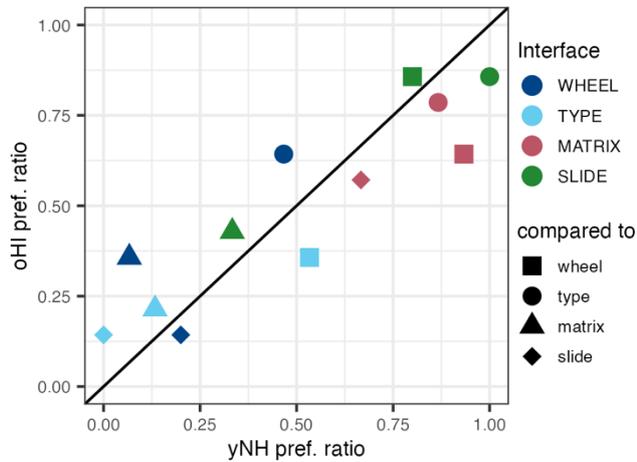

*Figure 7*. yNH vs. oHI preference ratios of the different interfaces. The ratio indicates how often an interface (color) was preferred over another interface (shape).

*Interface preferences*

To investigate how participants perceived the distinct UIs of the OLSA on a smartphone, we analyzed the comments participants provided using concept analysis. **Figure 8** visualizes the frequency of reported concepts across interfaces for both yNH and oHI. Examples for concepts, their explanation, and source comments can be found in **Table 3**. For a complete description of all concepts see **Supplementary Table 1**. The results are clustered to highlight the relevant concepts across interfaces. The concepts are, thus, in different order for yNH and oHI.

The analysis of the written and verbal comments of the participants revealed only slight differences across the two groups (yNH, oHI). The yNH provided more comments (mean = 7.13) than the oHI group (3.64). However, each participants provided at least one comment.

The type interface mainly received negative comments. For instance, it was reported by both yNH and oHI that the interface was cumbersome and difficult to handle. Similarly to the interface ranking, the wheel interface received rather negative comments by the yNH. The oHI noted that the wheel interface had a good size and was easy to use, but also that it was rather slow. The yNH also noted that it was slow, and indicated that it was annoying and cumbersome, among others. The oHI further noted that that they forgot the words while providing their response during a trial with the type and wheel interface.

The yNH group rated the matrix interface mostly positive. They perceived it as clear and easy, and noted positively that corrections were possible with this interface. For the slide interface, they instead negatively highlighted that no correction was possible, and errors occurred due to choosing a non-intended button. The oHI, in contrast, noted that the slide interface was easy and had good sizing. Though to a smaller amount, the oHI also pointed out that no correction was possible. In contrast to the slide interface, the oHI commented on the small size of the matrix interface.

Comparison of UIs for smartphone-based MSTs

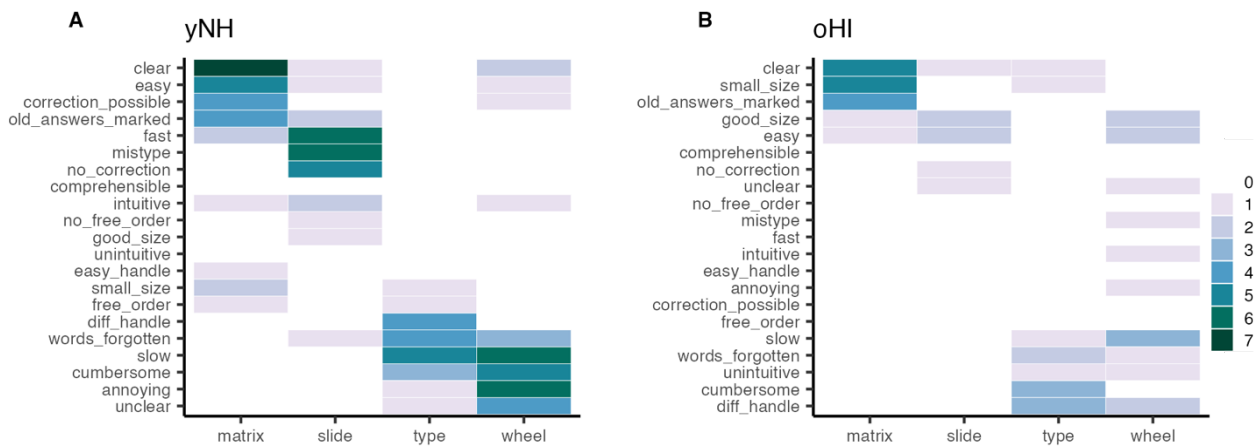

*Figure 8*. Prevalence of different concepts in the written and verbal comments of the participants. Concepts are clustered for visualization purposes. (A) displays the results for the yNH group; (B) the results for the oHI group.

*Table 3*. Exemplary concept names and explanations are shown, next to examples from provided comments (translated from German to English via DeepL for an unbiased and standardized translation). Underlined words highlight what caused a comment to be counted in the respective concept.

| Concept | Concept explanation | Example comment parts |
| --- | --- | --- |
| easy | Performing the tasks with this interface was easy or simple | - <u>Super easy to use</u>. Self-explanatory |
| Mistype | Mistyping occurred / wrong selection of a word | - similar, bad: quickly move on if <u>mistyped</u>; if last word tapped: immediately move on to next sentence<br>- I couldn't correct if I had <u>clicked the wrong way.</u> |
| no_correction | It was not possible to correct one's input | - In case of mistyping, is it possible to come back?<br>- I <u>couldn't correct</u> if I had clicked the wrong way. |
| Slow | It took time to complete the task with this interface, it was not fast | - time-intensive<br>- It <u>took quite a bit of time</u> to write it out, so I wasn't sure about the last few words. |
| words_forgotten | Sentence sound lost / Words forgotten / You had to remember the words for too long | - As I said, somewhat cumbersome and sometimes annoying, because you did not know in which direction the searched word was and <u>so forgot the other words.</u> An overview of the words available for selection would have been nice. |

## Discussion

The present study proved the feasibility of measuring the German Matrix sentence test (MST; OLSA) using a smartphone with household in-ear headphones for both yNH and oHI. We tested appropriate interfaces for a mobile implementation of the OLSA and compared them with respect to the following usability aspects: SRT consistency, required time, and user preferences. Finally, we discuss group effects (yNH vs. oHI) on interface usability and propose an interface for a mobile implementation.

Comparison of UIs for smartphone-based MSTs

*Feasibility of smartphone-based OLSA measurements*

The different interfaces resulted in different SRTs, completion times, and user preferences. The yNH results confirm that the OLSA could be tested in a valid way with all four distinct interfaces. This means, there are no interface-specific boundaries that hinder the assessment. The oHI group, conversely, demonstrates that potential age-dependent declines and/or unfamiliarity with smartphones can affect the precision of smartphone-based measurements.

Generally, we observed an increase in RMSEs with the oHI group. For all interfaces this increase was significant ($p < 0.05$). The increase for the matrix, slide, and wheel interface was, however, rather small. We therefore conclude that it is generally feasible to conduct the OLSA on a smartphone with inexpensive headphones, but there may be a slight SRT elevation with elderly hearing impaired individuals (RQ1). Further, we confirm that an interface effect is present within the oHI group (also evident from the interaction effects). In other words, the interfaces influence the consistency of the measurement for older adults with hearing impairments (RQ2). This highlights the necessity for including elderly in the development of smartphone-based MSTs and mobile health applications. Otherwise, applications may provide inaccurate results, or may not be easily usable and lead to frustration by the users (Kalimullah & Sushmitha, 2017).

*Group effects on the performance of the different interfaces*

*SRT consistency and preference*

The matrix interface resulted in the most consistent SRT scores, was the fastest OLSA interface, and was overall preferred both by yNH and oHI. For the yNH this could be expected, as the complete overview of the matrix is provided, the handling is fast, and smaller buttons are not expected to be an issue, due to no expected decline of motor skills. Conversely, it is surprising that the matrix interface was mostly positively received by the oHI and generally provided the most consistent results. Regardless, we do observe that one participant (outlier) was not able to achieve consistent results with the matrix interface and noted the small size of the interface. It could have been expected that the comparatively small button size and spacing would lead to larger difficulties with more participants of the oHI group, due to potentially reduced fine motor skills in the oHI group. However, several explanations exist that might explain the general SRT consistency and positive perception of the oHI group with the matrix interface.

First, the cognitive complexity of the matrix interface is smaller than with, for instance, the wheel and type interface. The Cognitive Complexity Theory (Kieras & Polson, 1985) in the context of UIs can be described as the number of different production rules to be learned. Production rules, in turn, are defined as IF (interface output) / THEN (user response) statements (Ziefle & Bay, 2005). To exemplify, following the playback of a sentence (IF), the user needs to perform different steps for task completion with the different interfaces. With the matrix interface, word buttons need to be selected (THEN). (IF) all buttons are selected, the next button needs to be selected for the next sentence to be played (THEN). With the slideshow interface, for instance, the last production rule is omitted, as the next sentence is played automatically after the last word is selected. The type interface, in contrast, requires further production rules (e.g. IF a keyboard appears, words need to be typed (THEN) and IF a word is suggested by the device, it can be selected (THEN)). The overall reduced cognitive complexity of the matrix interface may therefore have led, to a certain extent, to higher preferences of the oHI.

Comparison of UIs for smartphone-based MSTs

Second, a related explanation could be due to the enhanced error probability with increasing steps, or production rules. In that way, it could explain the higher SRT consistency of the matrix interface for most of the participants and the comparatively lower SRT consistency of the type interface. The probability of success can be modeled with: $p(success) = (1-p(failure))^{N(steps)}$ (Fisk et al., 2014). It follows, that the increasing number of steps with the type interface may have adversely affected the accuracy of the measurement.

Finally, fine motor skill declines may not be as pronounced in the present oHI group. Consequently, except for one participant, the oHI group may not have experienced that many difficulties in performing the task with the matrix interface, even though buttons and button spaces were comparatively small. Nonetheless, three participants had to be excluded as they were not capable of performing the tasks on a smartphone sufficiently without a pen. However, we have not assessed fine motor skills systematically. For future studies, it would be of interest to capture the exact impact of fine motor skills on UI preferences.

A preference difference between yNH and oHI individuals can be observed with the wheel interface (RQ2b). The wheel preference ratio was higher for oHI than yNH individuals in comparison to the type and matrix interface. For the slide interface there was no pronounced difference between the two groups. Considering that the matrix and type interface both displayed words in smaller font sizes and smaller button sizes than the wheel interface, this suggests a preference trend for larger font and button sizes with the oHI group.

*Completion time*

Besides SRT consistency and user preference, it is important for mobile measurements to be as fast as possible, whilst ensuring accurate results. Generally, users prefer faster measurements, and one might risk higher levels of inattentiveness and measurement abortion with slower measurements (Möckel et al., 2015). Longer completion times may also negatively impact the SRT consistency of the results. To exemplify, for the wheel and type interface both groups have indicated that they forgot words during the process of inputting the results.

As expected, the absolute completion time of the OLKISA was the shortest for both yNH and oHI, as it uses fewer trials and has a shorter sentence length than the OLSA. For the OLSA the matrix and slide interface were the fastest. Regardless, the oHI group took longer to perform the speech tests across all interfaces, and with the OLKISA. This is in line with research indicating that the cognitive decline with age and smartphone unfamiliarity leads to slower response time of elderly (Tsai et al., 2017). The reason for this is, however, that elderly choose their response more carefully to avoid errors. This change in response behavior can compensate for unfamiliarity with devices and can result in equally accurate results than younger users obtain (Starns & Ratcliff, 2010). This can be observed with the matrix interface, where SRT consistency was similar across groups, but the oHI took longer to complete the task.

The most striking group time difference can be observed with the type interface. Here, the time prolongation for completion of the oHI is larger than what would be expected from the observed time increases from the remaining interfaces. It is, thus, clear that the time prolongation goes beyond general potential speed reductions of the elderly, and instead demonstrates a group effect of this interface (RQ2b). An explanation could be that younger individuals are more used to writing text messages with their smartphones, than the elderly. Additionally, the tactile demand was highest in this interface, and therefore, the impairment by possible fine motor skill declines may be most noticeable here. Since the type interface took the longest, provided the least accurate SRT results, and was generally disliked by participants it can clearly be ruled out as a potential interface.

*Performance with spatial and fluctuating noise conditions*

Interestingly, we observed a bias in the ICRA5 condition with the slide interface, which we did not observe with the ICRA5 condition of the simplified matrix test OLKISA. Similarly, a slightly higher bias is present for the slide interface as compared to the matrix interface for the OLSA with ICRA1 (oHI). Consequently, the bias appears to stem specifically from the slide interface. We infer that higher accuracies would be achieved, if the slide interface would be adapted to address potential error sources resulting in the bias (see *Proposal of an adapted interface*), or the matrix interface would be used. A noteworthy observation is that the spatial and fluctuating noise conditions lead to a better SRT separability of the yNH and oHI group, in comparison to the S0N0 condition.

*Performance with the OLKISA (simplified matrix test)*

The OLKISA resulted in adequate consistencies, however, was not as consistent as the OLSA for the yNH group. This can be expected, as it is a reduced OLSA, both in trials (14 vs. 20) and in sentence length (3 vs. 5). Test-retest variability is also generally higher with simplified MSTs, than with MST. For instance, the test-retest value for the SRT50 with adults for the Italian MST and the simplified Italian MST are 0.6 dB and 1.2 dB, respectively (Puglisi et al., 2021). In our experiment, the average RMSE for yNH and oHI is 1.9. To a certain extent the RMSE value can therefore be explained by the general test-retest error. For an elderly target group with potential cognitive declines, the OLKISA can serve as a good alternative to the OLSA, if (1) none of the other interfaces seem plausible, and (2) the SRT consistency achieved with the OLKISA is considered as being sufficiently high.

### *Potential effects of smartphone-based measures and low-cost headphones*

Our study showed that cross-talk of headphones is an issue with non-professional audio equipment for conditions with a substantial interaural dissimilarity. This crosstalk can be due to "spatial audio processing" of the consumer electronic audio device, electric/electronic channel crosstalk, or due to an acoustic crosstalk path that all can be avoided under controlled laboratory conditions, but not for arbitrary consumer electronic devices. In our study, the monaural conditions with the OLSA (see Fig. 4, middle panel) appear to be affected by this effect: If the noise signal (intended to be played only to one ear) leaks over to the respective other headphone channel, normal hearing listeners can utilize the resulting binaural information to achieve a certain binaural release from masking which results in lowering the SRT in comparison to the reference value. While this effect can be seen for the monaural S0N90 condition for the yNH group, it is not evident for the oHI group due to three possible reasons: First, their expected SRT is closer to 0 dB which indicates that the presented speech signal will undergo the same crosstalk effect as the noise, thus making the mixed signal at both sides of the earphones more similar, thus excluding any binaural release from masking. Second, on average a much smaller binaural release from masking is present in hearing-impaired listeners, and third, the cross-talk may be below the (increased) threshold levels in the oHI group. For the monaural condition users with in-ear headphones could be instructed to remove the earplug for the ear to which no stimulus is presented (ear towards the noise direction). Alternatively, for over- and on-ear headphones, a cross-talk effect could be avoided by only measuring the spatial condition following an indication of hearing impairment via the S0N0 condition.

Comparison of UIs for smartphone-based MSTs

*Proposal of an adapted interface*

Not every participant could perform the task with the matrix interface, even though the interface resulted in accurate SRT scores for most participants. The oHI group also noted the small size of the words on the screen with this interface. This may be an issue with future participants and users with larger tactile difficulties than the participants employed in our study. Hence, another interface may be the preferable choice in the future. The slide and wheel interface both achieved adequate SRT consistencies. The slide interface, however, ranked higher in preference and was also faster to complete. In addition, one important outcome of the content analysis was that participants reported having accidentally pressed the wrong button, without the possibility to correct their mistake. Allowing participants to correct their mistake and to return to earlier parts of the sentence could decrease accidental error sources when performing the OLSA. As a result, the SRT deviations from the laboratory control session would be reduced. Additionally, the slide interface allows for an even further increase in font sizes. That way, a fast, consistent, and intuitive interface would be provided that avoids the small size when displaying the complete matrix. **Figure 9** shows the proposed adapted slide interface. We would therefore argue for the adapted slide interface to be implemented for mobile implementations of the OLSA (RQ3).

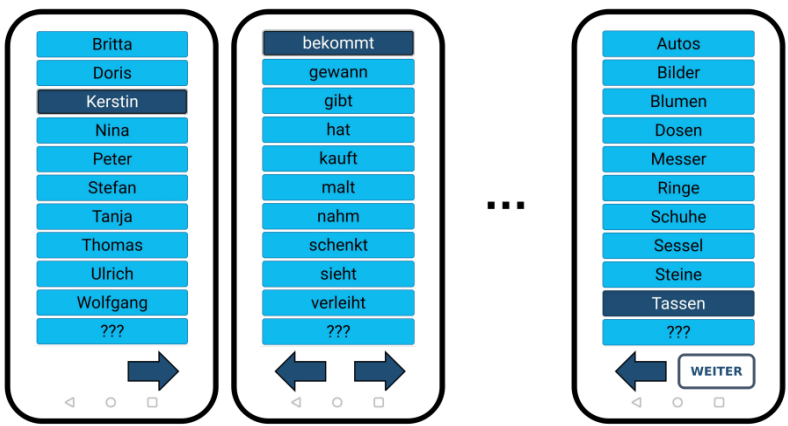

*Figure 9*. Proposed interface for performing the OLSA via a smartphone. Arrows allow to switch between the different "slides" or columns of the matrix. After a user has selected the understood words, pressing "Next" ("weiter" in German) will start the next sentence.

*Limitations and future research*

Even though the SRT is rather independent from the absolute presentation level and headphone transfer function, it is possible that some potential variation in SRT occurred due to loosening of the in-ear headphones. To estimate the potential effect for everyday life, investigations into the robustness of different headphone transfer functions (in-ear, on-ear, over-ear) against such mechanical variations and the resulting impact on the measured SRT would be of interest.

Furthermore, the tests were performed with calibrated equipment in a laboratory environment. It would be of interest to investigate how well the system would work in uncontrolled environments, with different uncalibrated setups (smartphones & headphones). The interfaces would not be expected to lead to different results in uncalibrated setups, but household headphones often have different frequency responses that boost certain frequencies and produce different overall sound levels. In addition, background noise may have an influence on measurement results. For that, it could be valuable to integrate a background noise monitoring system, as in Hussein et al (Hussein et al., 2016).

However, different sound levels are not expected to affect measurement acuity much, as the results of the matrix sentence tests (MST) are based on the SNR rather than on the absolute sound level and spectrum presented. Thus, it is less crucial to perform the measurements at the exact same noise level, and more important that the noise level is at least 20 dB higher than the PTA (Wardenga et al., 2018).

Finally, contrary to less efficient speech-in-noise test that provide higher test-retest variations in their results for a given amount of measurement time and hence cannot clearly detect a training effect, the matrix test exhibits a training effect of about 2 dB (Wagener, 2004). Hence, for clinical audiology purposes, training with two test lists of the MSTs is recommended. In our experiments, the experimenter conducted such a training session with a closed-set test version prior to the control and test session. Hence, participants were already familiar with the MST speech material when they did the measurements with the distinct interfaces. Being presented with the matrix interface may have been less surprising for the participants, than it could have been for naive participants. Displaying the interfaces to naive participants was not possible though, as the repeated measurements with the distinct interfaces would then have resulted in a training effect. For later implementation of the MST for an openly available web application where usually no training sessions can be performed, it would be important to control for the ongoing training effect, e.g., by monitoring the stability of the ongoing SRT estimate in real time.

*Conclusions*

Our study proves the general feasibility of measuring the matrix sentence test (OLSA) and the simplified matrix test (OLKISA) via a smartphone and in-ear headphones with both yNH and oHI individuals. For the four distinct interfaces overall group effects exist, next to group effects specific for distinct interfaces. The SRT's of the matrix interface were most consistent with the laboratory control for most of the participants, but not every participant could perform the tasks with this interface and the small size of the interface was noted. It is therefore less suitable for an elderly target group with potential tactile and visual impairments. The slide interface ranked second in terms of SRT consistency, preference, and completion time. At the same time, the participants provided suggestions for the slide interface (i.e., demanding the possibility of correcting the word selection), which could improve SRT consistency and stability. Therefore, the slide interface might eventually turn out to be a better alternative. The findings of this study should be applicable to the matrix sentence test in all available languages by translating the interfaces to the respective language.

**Funding**

This work was funded by the Deutsche Forschungsgemeinschaft (DFG, German Research Foundation) under Germanys Excellence Strategy −EXC 2177/1 – Project ID 390895286.

**Conflict of interest**

The authors declare that the research was conducted in the absence of any conflicts of interest.

Comparison of UIs for smartphone-based MSTs

Comparison of UIs for smartphone-based MSTs

**Supplementary Material**

*Supplementary Table 1.* The concept names and explanations are shown, next to examples from provided comments (translated from German to English via DeepL for an unbiased and standardized translation). Underlined words highlight what caused a comment to be counted in the respective concept in longer comments (with multiple concepts)

| Concept | Concept explanation | Example comment parts |
|---|---|---|
| Annoying | The interface had annoying properties | - Typing was super annoying<br>- As I said, somewhat cumbersome and sometimes annoying, because you did not know in which direction the searched word was and so forgot the other words. An overview of the words available for selection would have been nice. |
| clear | The layout was clear | - overview was positive<br>- Clearly designed and intuitive to use |
| Comprehensible | The interface was comprehensible/understandable | - Understandable<br>- It was comprehensible, but it took some time to find and select the right word, which often caused the sentence sound to be lost and the words could not be remembered. |
| correction_possible | It was possible to correct one's input | - a bit more confusing than B, but no problem if you get clicked wrong |
| Cumbersome | It was cumbersome to perform the task or the task was too complex | - As I said, somewhat cumbersome and sometimes annoying, because you did not know in which direction the searched word was and so forgot the other words. An overview of the words available for selection would have been nice. |
| diff_handle | The interface was difficult to handle | - it was difficult to write down the answer freely; unsure whether to capitalize the words or not<br>- Entering words into the text fields is quite difficult with this smartphone |
| easy_handle | The interface was easy to handle | - although small, but well selectable |
| easy | Performing the tasks with this interface was easy or simple | - Super easy to use. Self-explanatory |
| Fast | The interface was fast (timewise) | - Nice and simple, short and clear. …, fast, ... |
| free_order | One could freely choose the order of providing the answers | - Overview of all possibilities, any order of input possible |
| good_size | The size of the layout on the smartphone was good | - Font size big enough, scrolling no problem |

Comparison of UIs for smartphone-based MSTs

| | | |
|---|---|---|
| Intuitive | Performing the task with this layout was intuitive | - Clearly designed and intuitive to use<br>- Super easy to use. Self-explanatory |
| Old_answers_marked | Negative perception that old answers were still marked. | - Again, clicked fields remain highlighted, which can be a bit confusing |
| Mistype | Mistyping occurred / wrong selection of a word | - similar, bad: quickly move on if mistyped; if last word tapped: immediately move on to next sentence<br>-I couldn't correct if I had clicked the wrong way. |
| no_correction | It was not possible to correct one's input | - In case of mistyping, is it possible to come back? se fehlt auch noch Medical Physics und Uni Oldenburg, ist mir gerade noch aufgefallen<br>- I couldn't correct if I had clicked the wrong way. |
| no_free_order | One could not freely choose the order of providing the answer | - No free order |
| Slow | It took time to complete the in task with this interface, it was not fast | - time-intensive<br>- It took quite a bit of time to write it out, so I wasn't sure about the last few words. |
| small_size | The size of the layout on the smartphone was rather small | - Although small, but well selectable |
| Unclear | The layout was not clear | - ...,no overview of words,... |
| unintuitive | Performing the task with this layout was not intuitive | - unclear whether order of words is important; unclear why not also possible in one line |
| words_forgotten | Sentence sound lost / Words forgotten / You had to remember the words for too long | - As I said, somewhat cumbersome and sometimes annoying, because you did not know in which direction the searched word was and so forgot the other words. An overview of the words available for selection would have been nice. |